\documentclass[11pt,english]{article}
\usepackage{babel}
\usepackage[dvips]{graphics}
\usepackage{epsfig}
\usepackage{latexsym}
\usepackage{amscd,amsmath,amssymb}

\numberwithin{equation}{section}

\newcommand{\D}[1]{\operatorname{d#1}}
\newcommand{\Tr}[1]{\operatorname{Tr #1}}
\newcommand{\diag}[1]{\operatorname{diag}(#1)}
\newcommand{\tr}[1]{\operatorname{tr}#1}
\newcommand{\ad}[1]{\operatorname{ad}#1}

\newcommand{\sfrac}[2]{{\textstyle\frac{#1}{#2}}}
\newcommand{\ii}{\mbox{i}}
\newcommand{\e}{\mbox{e}}
\renewcommand{\=}{\ =\ }

\begin{document}

\begin{titlepage}

\begin{flushright}
hep-th/0211041\\
ITP--UH--24/02\\
\end{flushright}

\vskip 2.0cm

\begin{center}

{\LARGE\bf Noncommutative Instantons \\[12pt]
           via Dressing and Splitting Approaches}

\vspace*{0.4in}

{\Large 
Zal\'an Horv\'ath$^1$, \ Olaf Lechtenfeld$^2$ \ and \ Martin Wolf\,$^{2,3}$
}\\[0.4in]

  {\em $^1$Department of Theoretical Physics\\
       E\"otv\"os University\\
       1117, Budapest, Hungary\\[0.2in]

       $^2$Institut f\"ur Theoretische Physik\\
       Universit\"at Hannover\\
       30167, Hannover, Germany\\[0.2in]

       $^3$Institut f\"ur Theoretische Physik\\
       Technische Universit\"at Dresden\\
       01062, Dresden, Germany}

\vspace*{0.2in}

Email: zalanh@ludens.elte.hu, lechtenf, wolf@itp.uni-hannover.de

\end{center}

\vspace*{0.4in}

\begin{abstract}
\noindent
Almost all known instanton solutions in noncommutative Yang-Mills theory 
have been obtained in the modified ADHM scheme. In this paper we employ 
two alternative methods for the construction of the self-dual $U(2)$ BPST 
instanton on a noncommutative Euclidean four-dimensional space with 
self-dual noncommutativity tensor. Firstly, we use the method of dressing 
transformations, an iterative procedure for generating solutions from a 
given seed solution, and thereby generalize Belavin's and Zakharov's work 
to the noncommutative setup. Secondly, we relate the dressing approach 
with Ward's splitting method based on the twistor construction and rederive 
the solution in this context. It seems feasible to produce nonsingular
noncommutative multi-instantons with these techniques.
\end{abstract}

\end{titlepage}



\section{Introduction}

The idea of a noncommutative space-time is more than fifty years old 
\cite{Snyder:1947qz}. It offers a mild way to introduce nonlocality %
 into field theories without loosing too much control. Motiva%
ted by string theory \cite{Seiberg:1999vs} the investigation of non-Abelian g%
auge theories defined on noncommutative space-times took center stage %
 during the last couple of years. It is well known that 
 the dynamics of non-Abelian gauge fields involves nonperturbative field %
 configurations, like instantons and monopoles, in an essential way. 
Hence, before attempting to qu%
antize a gauge theory it is mandatory to study its classical solutions an%
d to characterize their moduli spaces. %

Nekrasov and Schwarz gave first examples of noncommutative instantons 
\cite{Nekrasov:1998ss}. They modified the Atiyah-Drinfeld-Hitchin-Manin (ADHM) 
method \cite{Atiyah:1978ri} and showed that (anti-self-dual) noncommutativity 
resolves the singularities of the instanton moduli space. Moreover, they 
discovered that the noncommutative Euclidean four-space $\mathbb{R}^4_\theta$ 
permits nonsingular Abelian instanton solutions, which do not survive in the 
commutative limit. Sinc%
e then, a lot of work has been devoted to this subject (see e.\,g.\ \cite%
{Braden:1999zp}-\cite{Franco-Sollova:2002nn} and references therein). For %
 reviews on this matter see \cite{Nekrasov:2000ih}-\cite{Nekrasov:2002kc}.

Most authors used the modified ADHM equations for the construction of inst%
antons on $\mathbb{R}^4_\theta$. In the present paper we advocate two com%
plementing methods termed the dressing approach and the twistor approach.
These proved to be successful in the commutative case: 
The twistor approach underlies the ADHM %
 scheme \cite{Atiyah:1978ri} and can produce generic $n$-instantons via a %
 sequence of Atiyah-Ward ans\"atze \cite{Atiyah:1977pw,WardWells}. %
 In particular, the 't Hooft $n$-instanton solution can be obtained in a %
very explicit way. The dressing approach \cite{zak} (see also 
\cite{Zakharov:pp}-\cite{Ward:vc}) produces this solution equally well 
\cite{Forgacs:1981su}.

In the noncommutative extension of $\mathbb{R}^4$ 
the classical fields are best represented
as (Lie-algebra valued) operators in an auxiliary two-oscillator Fock space.
The authors of \cite{Lechtenfeld:2001ie} employed the twistor approach to %
 construct $U(2)$ 't Hooft $n$-instantons on $\mathbb{R}^4_\theta$. Howev%
er, they were led to a gauge field which violates self-duality on an %
 $n$-dimensional subspace of the two-oscillator Fock space. This deficiency 
originated from the singular gauge choice pertinent to the 't Hooft solution
and was repaired by a suitable Murray-von Neumann transformation after a s%
pecific projection of the gauge potential. The proper noncommutative 't H%
ooft multi-instanton field strength was written down explicitly but its %
associated gauge potential could be given only implicitly. 
In order to get around these difficulties, Correa et al. \cite{Correa:2001wv} 
suggested to use the Belavin-Polyakov-Schwarz-Tyup%
kin (BPST) \cite{Belavin:fg} ansatz for constructing the noncommutative $%
U(2)$ one-instanton. In contrast to \cite{Lechtenfeld:2001ie} they did ob%
tain an explicit expression for the self-dual gauge potential but the rea%
lity of the gauge potential and field strength was lost.  

In this paper we concentrate on the one-instanton case for 
self-dual $\mathbb{R}^4_\theta$. Instead of trying %
 to generalize the BPST ansatz \cite{Belavin:fg} 
we choose the method of dressing trans%
formations and generalize the approach of Belavin and Zakharov 
\cite{Belavin:cz} to the noncommutative case. 
This eventually results in explicit 
expressions for a real gauge potential with self-dual field strength for 
the noncommutative $U(2)$ BPST instanton. Exploit%
ing the connection between the dressing method and Ward's splitting method we %
 rederive the same configuration by generalizing Crane's co%
nstruction \cite{Crane:im}. In fact, Crane's ansatz for the transition matr%
ix substitutes the Atiyah-Ward ansatz and leads directly to the nonsingul%
ar instanton configuration without the necessity of a singular gauge transf%
ormation. Its generalization may pave the way to nonsingular noncommutativ%
e multi-instantons.

The organization of the paper is as follows: First, we briefly discuss Y%
ang-Mills theory on commutative and noncommutative Euclidean four-dimensi%
onal space by introducing the basic notions and definitions. We then present %
the dressing method and illustrate it by constructing the noncommutative  
$U(2)$ BPST instanton solution for a self-dual noncommutativity tensor.  
In the last section we outline the connection be%
tween this method and the twistor approach and again provide the noncommu%
tative BPST instanton solution. An appendix briefly reviews the geometry of the
commutative twistor space.


\section{Yang-Mills theory on commutative $\mathbb{R}^4$ and noncommutati%
ve $\mathbb{R}^4_{\theta}$}

\paragraph{Commutative Yang-Mills theory.} We consider the Euclidean four%
-dimensional space $\mathbb{R}^4$ with the canonical metric $\delta_{\mu
\nu}$. Furthermore we specify to a principal bundle of the form $P=\mathbb{R%
}^4\times U(2)$ with a connection ${A}={A}_{\mu}\D{x^%
\mu}$ and the Yang-Mills curvature ${F}=\D{}{A}+%
{A}\wedge{A}$. In components the latter equation reads $%
{F}_{\mu\nu}=\partial_{\mu}{A}_{\nu}-\partial_{\nu}{A}_%
{\mu}+[{A}_{\mu},{A}_{\nu}]$. Here, $\partial_{\mu}$ denot%
es the partial derivative with respect to $x^\mu$, and Greek 
indices always run from $1$ to $4$.

The self-dual Yang-Mills (SDYM) equations take the form
\begin{equation}\label{sdym}
 {F}\=*{F}\qquad\Longleftrightarrow\qquad{F}_{\mu\nu%
}\=\sfrac{1}{2}\epsilon_{\mu\nu\rho\sigma}{F}^{\rho\sigma}\ ,
\end{equation}
where `$*$' denotes the Hodge star and $\epsilon_{\mu\nu\rho\sigma}$ is t%
he Levi-Civita symbol on $\mathbb{R}^4$ with $\epsilon_{1234}=1$. Solut%
ions of (\ref{sdym}) having finite Yang-Mills action are called instanton%
s. Their action is given by
\begin{equation}\label{action}
 S \= -\frac{1}{g^2_{\text{\tiny YM}}}\int\tr {F}\wedge{*%
{F}}\=\frac{8\pi^2}{g^2_{\text{\tiny YM}}}Q\ ,
\end{equation}
where $Q$ is the topological charge
\begin{equation}
 Q \= -\frac{1}{8\pi^2}\int\tr {F}\wedge{F}\ .
\end{equation}
Here, `$\tr$' is the trace over the $\mathfrak{u}(2)$ gauge algebra and $g%
_{\text{\tiny YM}}$ the Yang-Mills coupling hidden in the definition of t%
he Lie algebra components of ${A}$ and ${F}$.

Introducing complex coordinates\footnote{This particular choice and its g%
eometrical meaning will be clarified in section four and in the appendix.%
}
\begin{equation}\label{cc}
 z^1=x^1+\ii x^2, \quad\bar{z}^1=x^1-\ii x^2, \quad 
 z^2=x^3-\ii x^4, \quad\bar{z}^2=x^3+\ii x^4
\end{equation}
and defining %
\begin{subequations}
\begin{equation}
 {A}_{z^1}\=\sfrac{1}{2}({A}_1-\ii{A}_2)\qquad\text%
{and}\qquad%
 {A}_{z^2}\=\sfrac{1}{2}({A}_3+\ii{A}_4)
\end{equation}
as well as
\begin{equation}
 {A}_{\bar{z}^1}\=\sfrac{1}{2}({A}_1+\ii{A}_2)\qquad
\text{and}\qquad%
 {A}_{\bar{z}^2}\=\sfrac{1}{2}({A}_3-\ii{A}_4)\ ,
\end{equation}
\end{subequations}
we can rewrite the SDYM equations (\ref{sdym}) in the form
\begin{equation}\label{eom}
 [D_{z^1},D_{z^2}]=0\ ,\quad
 [D_{\bar{z}^1},D_{\bar{z}^2}]=0\ ,\quad
 [D_{z^1},D_{\bar{z}^1}]+[D_{z^2},D_{\bar{z}^2}]=0\ ,
\end{equation}
with $D_{z^a}=\partial_{z^a}+{A}_{z^a}$ 
and $D_{\bar{z}^a}=\partial_{\bar{z}^a}+{A}_{\bar{z}^a}$, 
respectively for $a=1,2$.

\paragraph{Noncommutative Yang-Mills theory.} 
A noncommutative extension of $\mathbb{R}^4$ is defined via deforming the
ring of functions on it. More precisely, the pointwise product between 
functions gets replaced with the Moyal star product which is defined by
\begin{equation}\label{sp}
 (f\star g)(x)\ :=\ f(x)\,\exp\left\{\sfrac{\ii}{2}\overleftarrow{\partial_{\mu
}}\theta^{\mu\nu}\overrightarrow{\partial_{\nu}}\right\}\,g(x)\ ,
\end{equation}
where $f,g\in C^{\infty}(\mathbb{R}^{4},\mathbb{C})$ and $\theta^{\mu\nu}$ is 
chosen to be a constant antisymmetric tensor. Equation (\ref{sp}) implies that
\begin{equation}
 [x^{\mu},x^{\nu}]_{\star}\ :=\ x^{\mu}\star x^{\nu}-x^{\nu}\star x^{\mu}\=%
\ii\theta^{\mu\nu}\ .
\end{equation}%
In this paper we restrict $\theta^{\mu\nu}$ to be self-dual with
\begin{equation}\label{cot}
 \theta^{12}=-\theta^{21}=\theta^{34}=-\theta^{43}=:\theta>0
\end{equation}
and all other components being identically zero. The action (\ref{action}) and
the SDYM equations (\ref{sdym}) are formally unchanged, but the ordi%
nary product needs to be replaced by the star product, e.\,g.
\begin{equation}
 {F}_{\mu\nu}\=\partial_{\mu}{A}_{\nu}-\partial_{\nu}%
{A}_{\mu}+{A}_{\mu}\star{A}_{\nu}-{A}_{\nu}\star{A}_{\mu}\ .
\end{equation}

It is well known that the nonlocality of the star product can be cumbersome
for explicit calculations. It is therefore convenient to pass %
to the operator formalism via the Weyl correspondence given by
\begin{subequations}
\begin{eqnarray}
 \tilde{f}(k)\ \mapsto\ \hat{f}(\hat{x}) &=& \frac{1}{(2\pi)^4}\int\D{ ^4k}%
\,\tilde{f}(k)\,\e^{\ii k\hat{x}}\ ,\\
 \hat{f}(\hat{x})\ \mapsto\ \tilde{f}(k) &=& (2\pi\theta)^2\,
\Tr{}\bigl\{\e^{-\ii k\hat{x}}\hat{f}(\hat{x})\bigr\}\ ,
\end{eqnarray}
\end{subequations}
where `$\Tr{}$' denotes the trace over the operator representation of the %
 noncommutative algebra and $\tilde{f}(k)$ is the Fourier transform of $f%
(x)$, i.\,e.
\begin{equation}
 f(x)\ \mapsto\ \tilde{f}(k)\=\int\D{ ^4x}\,f(x)\,\e^{-\ii kx}\ .
\end{equation}
In these equations $kx$ is shorthand notation for $k_{\mu}x^{\mu}$. 
Important relations are
\begin{equation}
 f\star g\ \mapsto\ \hat{f}\,\hat{g}\qquad\text{and}\qquad\int\D{ ^4x}\,f\=(2%
\pi\theta)^2\Tr{}\hat{f}\ ,
\end{equation}
whenever both sides of the latter equation exist.
Thus, we obtain operator-valued coordinates $\hat{x}^\mu$ satisfying the %
commutation relations $[\hat{x}^\mu,\hat{x}^\nu]=\ii\theta^{\mu\nu}$ which
define the noncommutative Euclidean four-dimensional space denoted by %
 $\mathbb{R}^4_\theta$. 

In complex coordinates $(\ref{cc})$ our choice of %
 $\theta^{\mu\nu}$ (\ref{cot}) implies
\begin{equation}\label{cr}
 [\hat{z}^1,\hat{\bar{z}}^1]\=2\theta\qquad\text{and}\qquad[\hat{z}^2,\hat%
{\bar{z}}^2]\=-2\theta
\end{equation}
with all other commutators being equal to zero.
Coordinate derivatives turn into inner derivations of the operator algebra, 
i.\,e.
\begin{subequations}
\begin{equation}\label{der1}
 \hat{\partial}_{z^1}\hat{f}       \= -\frac{1}{2\theta}\ad{}(\hat{\bar{%
z}}^1)(\hat{f})\qquad\text{and}\qquad
 \hat{\partial}_{\bar{z}^1}\hat{f} \= \frac{1}{2\theta}\ad{}(\hat{z}^1)(%
\hat{f})
\end{equation}
as well as
\begin{equation}\label{der2}
 \hat{\partial}_{z^2}\hat{f}       \= \frac{1}{2\theta}\ad{}(\hat{\bar{z%
}}^2)(\hat{f})\qquad\text{and}\qquad
 \hat{\partial}_{\bar{z}^2}\hat{f} \= -\frac{1}{2\theta}\ad{}(\hat{z}^2)%
(\hat{f})\ .
\end{equation}
\end{subequations}
Since the commutation relations (\ref{cr}) identify our operator algebra
with a pair of Heisenberg algebras it can obviously be represented on the
two-oscillator Fock space $\mathcal{H}=\bigoplus_{n_1,n_2}\mathbb{C}\,|n%
_1,n_2\rangle$ with
\begin{equation}
 |n_1,n_2\rangle \= \frac{1}{\sqrt{n_1!n_2!(2\theta)^{n_1+n_2}}}\,(\hat{\bar%
{z}}^1)^{n_1}(\hat{z}^2)^{n_2}|0,0\rangle\ .
\end{equation}
Hence, coordinates and fields on $\mathbb{R}^4_\theta$ both correspond to 
operators acting on $\mathcal{H}$. 
Sitting in the adjoint representation of $\mathfrak{u}(2)$
the components ${\hat{A}}_\mu$ and $%
{\hat{F}}_{\mu\nu}$ act from the left in the space 
$\mathcal{H}\otimes\mathbb{C}^2\cong\mathcal{H}\oplus\mathcal{H}$, 
which carries %
a fundamental representation of the group $U(2)$. In the operator formula%
tion the action (\ref{action}) reads
\begin{equation}
 S\=-\sfrac{1}{2}\Bigl(\frac{2\pi\theta}{g_{\text{\tiny YM}}}\Bigr)^2
 \Tr{}\bigl\{\tr{}{\hat{F}}_{\mu\nu}{\hat{F}}^{\mu\nu}\bigr\}\ ,
\end{equation}
and the SDYM equations (\ref{eom}) retain their familiar form
\begin{equation}\label{eom-2}
 \hat{{F}}_{z^1z^2}=0\ ,\quad
 \hat{{F}}_{\bar{z}^1\bar{z}^2}=0\ ,\quad
 \hat{{F}}_{z^1\bar{z}^1}+\hat{{F}}_{z^2\bar{z}^2}=0\ .
\end{equation}
In order to streamline our notation we will from now on omit the hats %
over the operators.


\section{Instantons from the dressing approach}

\paragraph{Linear system.} The key observation is that the SDYM equations %
 (\ref{eom-2}) can be obtained as the integrability conditions of the lin%
ear system (already in the noncommutative setup)
\begin{equation}\label{ls}
 (D_{\bar{z}^1}-\lambda D_{z^2})\psi \=0\ , \qquad (D_{\bar{z}^2}+\lambda %
 D_{z^1})\psi \= 0
 \qquad\text{and}\qquad\partial_{\bar{\lambda}}\psi\=0,
\end{equation}
where $\lambda\in\mathbb{C}\,\cup\,\{\infty\}\cong\mathbb{CP}^1$ is calle%
d the spectral parameter and the $2{\times}2$ matrix $\psi$ depends on $(x%
^1, x^2, x^3, x^4, \lambda)$ (or equivalently on $(z^1,\bar{z}^1,z^2,\bar%
{z}^2,\lambda)$). Hence, it lives on a space $\mathcal{P}=\mathbb{R}%
^4_\theta\times\mathbb{CP}^1$ known as the twistor space. Since this syst%
em is overdetermined there are integrability conditions which turn o%
ut to be exactly the SDYM equations (\ref{eom-2}).

Following Belavin and Zakharov \cite{Belavin:cz} we impose a reality co%
ndition on $\psi$:
\begin{equation}\label{rc}
 [\psi(x,-\bar{\lambda}^{-1})]^\dagger\=\psi^{-1}(x,\lambda)\ ,
\end{equation}
with $x\in\mathbb{R}^4_\theta$. This restriction ensures that the %
 gauge potential ${A}_\mu$ is anti-Hermitian, i.\,e.\ ${A}^%
\dagger_\mu=-{A}_\mu$, as is required by the gauge group $U(2)$.

The linear system (\ref{ls}) can be rewritten as
\begin{subequations}\label{ls1}
\begin{eqnarray}
 \psi(\partial_{\bar{z}^1}-\lambda\partial_{z^2})\psi^\dagger &=&%
 {A}_{\bar{z}^1}-\lambda {A}_{z^2}\ ,\\
 \psi(\partial_{\bar{z}^2}+\lambda\partial_{z^1})\psi^\dagger &=&%
 {A}_{\bar{z}^2}+\lambda {A}_{z^1}\ ,
\end{eqnarray}
\end{subequations}
whereby $\psi^\dagger$ is shorthand notation for $[\psi(x,-\bar{\lambda}^%
{-1})]^\dagger$. The task is to find the auxiliary field $\psi$, sinc%
e the gauge potential and, hence, the curvature then follow immediately from %
the equations above.%

\paragraph{Dressing method.} 
In the commutative setup Belavin and Zakharov constructed \cite{
Belavin:cz} solutions to the SDYM equations (\ref{sdym}) 
by using the method of dressing transformations, which is a recursive pr%
ocedure for generating solutions from a given seed solution. After having
derived the one-instanton BPST solution they used it as the seed %
solution to give a recursion relation for the construction of the $n$-ins%
tanton configuration. The extension of the dressing method to the noncommutat%
ive case is readily accomplished (cf. \cite{Lechtenfeld:2001aw}, \cite{
Lechtenfeld:2001gf} and \cite{Wolf:2002jw}). We now recall this extension
and generalize the Belavin-Zakharov ansatz to the noncommutative realm.

Let $\psi$ be a given solution of the linear system (\ref{ls1}). To gener%
ate a new solution out of $\psi$ we multiply the so-called dressing facto%
r $\chi$ on the left, i.\,e.\ we consider
\begin{equation}
 \tilde{\psi}(z^1,\bar{z}^1,z^2,\bar{z}^2,\lambda)\=\chi(z^1,\bar{z}^1,z%
^2,\bar{z}^2,\lambda)\,\psi(z^1,\bar{z}^1,z^2,\bar{z}^2,\lambda)\ .
\end{equation}
The dressing factor $\chi$ is assumed to be a global meromorphic operator%
-valued\footnote{When we say operator-valued function we imply a $\{z^a\}%
$ and $\{\bar{z}^a\}$ dependence.} function in the spectral parameter 
$\lambda\in\mathbb{CP}^1$, which implies the expansion
\begin{equation}\label{ansatzchi}
 \chi \= \lambda R_{-1}+R_0+\sum_{i=1}^r\frac{R_i}{(\mu_i\lambda+\nu_i%
)}
\end{equation}
with $\mu_i,\,\nu_i\in\mathbb{C}$ and the coefficients $R_{-1}$, $R_0$ an%
d $R_i$ depending on $\{z^a\}$ and $\{\bar{z}^a\}$ but not on the spectral p%
arameter $\lambda$. We have restricted ourselves to first-order pole%
s, since one can generate higher-order poles by a successive multipl%
ication of such expressions. Thus, $\chi$ has finitely many pol%
es which are located at $\lambda_\infty=\infty$ and $\lambda_i=-\nu_i%
/\mu_i$ with $\mu_i\neq 0$ for $i=1,\ldots,r$ .

Since $\tilde{\psi}$ is supposed to be a new solution we can write
\begin{subequations}
\begin{eqnarray}
 \tilde{\psi}(\partial_{\bar{z}^1}-\lambda\partial_{z^2})\tilde{\psi}^\dagger%
             &=& \tilde{{A}}_{\bar{z}^1}-\lambda \tilde{%
{A}}_{z^2}\ ,\\
 \tilde{\psi}(\partial_{\bar{z}^2}+\lambda\partial_{z^1})\tilde{\psi}^\dagger%
             &=& \tilde{{A}}_{\bar{z}^2}+\lambda \tilde{%
{A}}_{z^1}\ .
\end{eqnarray}%
\end{subequations}
A short calculation shows that
\begin{subequations}\label{efchi}
\begin{eqnarray}
 \chi(D_{\bar{z}^1}-\lambda D_{z^2})\chi^\dagger &=& \tilde{{A}%
}_{\bar{z}^1}-\lambda \tilde{{A}}_{z^2}\ ,\\
 \chi(D_{\bar{z}^2}+\lambda D_{z^1})\chi^\dagger &=& \tilde{{A}%
}_{\bar{z}^2}+\lambda \tilde{{A}}_{z^1}\ ,
\end{eqnarray}%
\end{subequations}
where $D_\mu=\partial_\mu+{A}_\mu$ is the covariant derivative in the 
background of the old gauge potential ${A}_\mu$ determined through $\psi$. 
Since the gauge potential $\tilde{{A}}_\mu$ is $\lambda$-independent
the left hand side of (\ref{efchi}) is at most linear in $\lambda$. 
The ansatz (\ref{ansatzchi}) for $\chi$, however, contains finitely many 
poles in the spectral parameter. Therefore, all proper residues must vanish. 
This requirement yields differential equations for the coefficients 
$R_{-1}$, $R_0$ and $R_i$ for $i=1,\ldots,r$. After solving these
equations one obtains a new solution $\tilde{{A}}_\mu$ of the noncommutative 
SDYM equations. Finally, this procedure may be iterated 
to get new solutions from old ones.

\paragraph{Ansatz and noncommutative BPST instanton.} 
Let us obtain a one-instanton solution by way of dressing. 
The trivial solution of 
(\ref{ls1}) is $\psi=1$ and ${A}_\mu=0$. We take this solution a%
s the seed solution and choose the dressing factor of the form \cite{
Belavin:cz}\footnote{ Here we have renamed the d%
ressing factor $\chi$ and called it $\psi$ in order to be conform with th%
e literature.}
\begin{equation}\label{ansatzpsi}
 \psi(x,\lambda) \= G\,\Bigl(1+2H+\lambda S^\dagger+\frac{1}{\lambda}S\Bigr)\ ,
\end{equation}
where all $\lambda$ dependence is made explicit and 
$G$ and $H$ are taken to be Hermitian diagonal $2{\times}2$ matrix
functions of $\{z^a,\bar{z}^a\}$, i.\,e.
\begin{equation} \label{diag}
G\ =:\ \diag{g_-,g_+} \qquad\text{and}\qquad H\ =:\ \diag{h_-,h_+}\ .
\end{equation}
{}From equation (\ref{ansatzpsi}) we obtain
\begin{equation}
 [\psi(x,-\bar{\lambda}^{-1})]^\dagger\=\Bigl(1+2H-\lambda S^\dagger-
\frac{1}{\lambda}S\Bigr)\,G\ .
\end{equation}
In order to simplify further calculations we put some restrictions on $H$ %
 and $G$, namely we require that 
\begin{equation}
[G,H]\=[G,S]\=[H,S]\=0\ .\label{commute} 
\end{equation}
Then the reality condition (\ref{rc}) implies
\begin{subequations}
\begin{equation}
 S^2\=0 \qquad\text{and}\qquad  (S^\dagger)^2\=0 \label{cog2}
\end{equation}
as well as
\begin{equation}
 G^2(1{+}2H)^2 \= 1+G^2\{S,S^\dagger\}\ . \label{cog}
\end{equation}
\end{subequations}
The $\lambda$ dependence of the differential equations (\ref{ls1}) leads to
\begin{subequations}
\begin{equation}
 G\bigl\{S^\dagger\partial_{z^a}S^\dagger\bigr\}G\=0
 \qquad\text{for}\quad a=1,2 \label{cpde}
\end{equation}
as well as
\begin{equation}
 G\bigl\{(1{+}2H)\partial_{z^a}S^\dagger-S^\dagger\partial_{z^a}(1{+}2H)
 +\epsilon_{ab}S^\dagger\partial_{\bar{z}^b}S^\dagger\bigr\}G\=0
 \qquad\text{for}\quad a=1,2\ ,\label{cpde2}
\end{equation}
\end{subequations}
where the nilpotency of $S^\dagger$ has been used.
We shall find matrix functions $H$ and $S$ for which the brackets above 
will vanish by themselves.

In order to construct a $2{\times}2$ matrix~$S$ satisfying 
(\ref{cog2}) and (\ref{cpde}) (which do not involve $H$),
we consider the two vectors
\begin{equation}
 T_1\=\binom{v_1}{-v_2}\qquad\text{and}\qquad 
 T_2\=\binom{{v}^{\dagger}_2}{{v}^{\dagger}_1}
\end{equation}
built from functions $v_1$ and $v_2$ and introduce 
some real function $K=K(r^2)$ which depends only on the combination
\begin{equation}
 r^2\ :=\ \bar{z}^1z^1+\bar{z}^2z^2\ .
\end{equation}
With these ingredients we parametrize $S^\dagger$ as follows,
\begin{equation}
 S^\dagger\=T_1\frac{1}{K}T^\dagger_2\=\begin{pmatrix} 
 v_1\frac{1}{K}v_2  &  v_1\frac{1}{K}v_1 \\
 -v_2\frac{1}{K}v_2 & -v_2\frac{1}{K}v_1 \end{pmatrix}\ ,
\end{equation}
and expect conditions on $v_1$, $v_2$, and $K$.
Due to the nilpotency property (\ref{cog2}) of $S$ and $S^\dagger$ we %
 find $[v_1,v_2]=0$ and hence, $T^\dagger_1T_2=T^\dagger_2T_1=0$. T%
he condition (\ref{cpde}) then tells us that $v_1$ and $v_2$ are anti-holomorp%
hic functions depending on $\bar{z}^1$ and $\bar{z}^2$ only. A simple choice i%
s $v_1=\bar{z}^1$ and $v_2=\bar{z}^2$, which specifies 
\begin{equation}
 S^\dagger \= \begin{pmatrix} \bar{z}^1\frac{1}{K}\bar{z}^2  & \bar{z}^1%
 \frac{1}{K}\bar{z}^1 \\%
                     -\bar{z}^2\frac{1}{K}\bar{z}^2 & -\bar{z}^2\frac{1}{%
 K}\bar{z}^1 \end{pmatrix}\ .
\end{equation}

The commutation property (\ref{commute}) of $G$ and $H$ with this form of $S$ 
can be achieved by choosing $g_\pm$ and $h_\pm$ in (\ref{diag})
to depend only on $r^2$ and by relating them via
\begin{equation}
 g_\pm(r^2)\=g(r^2\pm 2\theta) \qquad\text{and}\qquad
 h_\pm(r^2)\=h(r^2\pm 2\theta)
\end{equation}
to yet unknown scalar functions $g$ and $h$.
It is useful to define the notation $f_\pm(r^2):=f(r^2\pm 2\theta)$
for arbitrary functions~$f$.

Next we address equations (\ref{cpde2}). 
Remembering that partial derivatives are given by the inner derivations
(\ref{der1}) and (\ref{der2}) we reformulate:
\begin{subequations}
\begin{eqnarray}
 G\bigl\{(1{+}2H)\,\bar{z}^1S^\dagger-S^\dagger\bar{z}^1(1{+}2H)
         +S^\dagger z^2S^\dagger\bigr\}G &=& 0\ ,\label{eom1}\\
 G\bigl\{(1{+}2H)\,\bar{z}^2S^\dagger-S^\dagger\bar{z}^2(1{+}2H)
         -S^\dagger z^1S^\dagger\bigr\}G &=& 0\ .\label{eom2}
\end{eqnarray}
\end{subequations} 
Demanding that the brackets vanish gives us two relations for the
functions $h$ and $K$. We claim that these are fulfilled for
\begin{equation} \label{coT}
 h(r^2)\=-\sfrac{1}{2}\Bigl(1+\frac{K(r^2)+2C}{K(r^2)}\Bigr) 
 \qquad\text{and}\qquad K(r^2+2\theta)\=K(r^2)+2\theta 
\end{equation}
with $C\in\mathbb{R}$ being some (almost) arbitrary constant.
The proof of this assertion is rather straightforward and we therefore 
refrain from presenting it here.
The general solution to the functional equation in (\ref{coT}) is
\begin{equation}
 K(r^2) \= r^2 + \tilde{K}(r^2)
\end{equation}
with an arbitrary $2\theta$-periodic function $\tilde{K}$.
By imposing a smooth $\theta\to0$ limit we force $\tilde{K}$ to be constant,
i.\,e.
\begin{equation}\label{sfT}
 K(r^2) \= r^2 - 2C + 2\Lambda^2
\end{equation}
with some new real constant $\Lambda^2\ge0$.\footnote{
The positivity of $\Lambda^2$ will follow from the hermiticity of $g$.}
We are going to see that $\Lambda$ corresponds to the size of the instanton.
The corresponding $h$ reads
\begin{equation}
 h(r^2)\=-\frac{r^2+2\Lambda^2-C}{r^2+2\Lambda^2-2C}\ .
\end{equation}
A particularly simple choice is $C=0$ which results in $h=-1$.
More generally, we should only exclude the values $C=\Lambda^2+n\theta$
for $n\in\mathbb{Z}_+$, where $H$ becomes singular.

It remains to impose the condition (\ref{cog}) which serves to determine
the function $g$. With the help of 
\begin{equation}
 \{S,S^\dagger\}\=\begin{pmatrix} \frac{(r^2)_-^2}{K_-^2} & 0 \\ 
                                 0 & \frac{(r^2)_+^2}{K_+^2} \end{pmatrix}
\end{equation}
equation (\ref{cog}) reduces to
\begin{equation}
 g^2(1{+}2h)^2 \= 1+g^2\frac{(r^2)^2}{K^2} 
\end{equation}
which yields
\begin{equation} \label{gsol}
 g(r^2)\=\pm\frac{1}{2\Lambda}\frac{r^2+2\Lambda^2-2C}{\sqrt{r^2+\Lambda^2}}\ .
\end{equation}

During this computation we tacitly assumed that all quantities are well 
behaved such that we could legally perform all necessary operations. 
However, this need not always be the case.
As a prime example, $(r^2)_-^{-1}\equiv(r^2{-}2\theta)^{-1}$ is ill behaved 
when acting on the ground state $|0,0\rangle $ of $\mathcal{H}$. 
Luckily, the potentially dangerous $K_-^{-1}$ is regulated by $\Lambda^2$.
Only the limit $\Lambda^2\to0$ is apparently singular. 
The results above may be recombined to compose the final expression for $\psi$:
\begin{equation}\label{solpsi}
 \psi\=\frac{1}{2\Lambda}
       \begin{pmatrix} \frac{1}{\sqrt{r^2+\Lambda^2-2\theta}} & 0\\%
                                                            0 &  \frac{1}%
{\sqrt{r^2+\Lambda^2+2\theta}}
       \end{pmatrix}
       \begin{pmatrix} {\scriptstyle r^2+2\Lambda^2-2\theta-\lambda \bar{%
z}^1\bar{z}^2-}\frac{z^1z^2}{\lambda}%
         & {\scriptstyle -\lambda(\bar{z}^1)^2+}\frac{(z^2)^2}{\lambda}\\%

           {\scriptstyle \lambda(\bar{z}^2)^2-}\frac{(z^1)^2}{\lambda}%
         & {\scriptstyle r^2+2\Lambda^2+2\theta+\lambda\bar{z}^1\bar{z}^2%
+}\frac{z^1z^2}{\lambda}
       \end{pmatrix}\ ,
\end{equation}
where we have chosen the negative sign in (\ref{gsol}). 
The $\theta$ dependence of this noncommutative solution is very simple.
It is also quite remarkable that it differs from the commutative solution 
(as given by Belavin and Zakharov \cite{Belavin:cz}) only in a few spots.
Moreover, the final expression is independent of the parameter $C$, 
in accordance with the latter's interpretation as a regulator of intermediate
singularities in the course of the calculation.

\paragraph{Connection and curvature.} 
We are now in the position to construct the ga%
uge potential using our solution (\ref{solpsi}). By looking at the terms of %
(\ref{ls1}) which are linear in $\lambda$ we can compute ${A}_{z^%
1}$ and ${A}_{z^2}$. The calculation is kind of t%
edious but straightforward. We therefore omit the %
explicit derivation and give only the results
\begin{subequations}\label{gauge}
\begin{equation}
 {A}_{z^1}\=\begin{pmatrix} -\frac{\bar{z}^1}{2\theta}\biggl(\sqrt{
\frac{r^2+\Lambda^2-2\theta}{r^2+\Lambda^2}}-1\biggr) & 0\\
                   -{\scriptstyle \bar{z}^2}\frac{1}{\sqrt{r^2+\Lambda^2}%
\sqrt{r^2+\Lambda^2-2\theta}} &
                   -\frac{\bar{z}^1}{2\theta}\biggl(\sqrt{\frac{r^2+
\Lambda^2+4\theta}{r^2+\Lambda^2+2\theta}}-1\biggr) \\
                   \end{pmatrix}
\end{equation}
and
\begin{equation}
 {A}_{z^2}\=\begin{pmatrix} \biggl(\sqrt{\frac{r^2+\Lambda^2-2\theta}{
r^2+\Lambda^2}}-1\biggr)\frac{\bar{z}^2}{2\theta} &
                   -\frac{1}{\sqrt{r^2+\Lambda^2}\sqrt{r^2+\Lambda^2-2
\theta}}{\scriptstyle \bar{z}^1}\\
                   0 & \biggl(\sqrt{\frac{r^2+\Lambda^2+4\theta}{r^2+\Lambda^2
+2\theta}}-1\biggr)\frac{\bar{z}^2}{2\theta}
                   \end{pmatrix}\ ,
\end{equation}
\end{subequations}
which in the commutative limit $\theta\to 0$ coincide with the Belavin-Za%
kharov solution \cite{Belavin:cz}. 
The remaining components are ${A}_{\bar{z}^a}=-{A}^\dagger_{z^a}$ for $a=1,2$.

Furthermore, we compute the nonvanishing components of the Yang-Mills curvature
using $F_{z^a\bar{z}^b}=\partial_{z^a}A_{\bar{z}^b}%
-\partial_{\bar{z}^b}A_{z^a}+[A_{z^a},A_{\bar{z}^b}]$
for $a,b=1,2$. 
Again, this task is lengthy but not difficult. 
Ultimately we find the self-dual configuration
\begin{subequations}\label{curv}
\begin{equation}\label{curv1}
 {F}_{z^1\bar{z}^1}\=-{F}_{z^2\bar{z}^2}\=
                           \begin{pmatrix} -\frac{\Lambda^2}{(r^2+\Lambda%
^2)(r^2+\Lambda^2-2\theta)} & 0\\
                           0 & \frac{\Lambda^2}{(r^2+\Lambda^2-2\theta)(r%
^2+\Lambda^2+2\theta)} \end{pmatrix}
\end{equation}
and
\begin{equation}\label{curv2}
 {F}_{z^1\bar{z}^2}\={F}^\dagger_{z^2\bar{z}^1}\=
                           \begin{pmatrix} 0 & 0 \\ %
                           \frac{2\Lambda^2}{(r^2+\Lambda^2)\sqrt{r^2+
\Lambda^2-2\theta}\sqrt{r^2+\Lambda^2+2\theta}} & 0%
                           \end{pmatrix}.
\end{equation}
\end{subequations}
Comparing this solution, constructed by the dressing method, 
with the one obtained via the ADHM approach \cite{Furuuchi:2001dx} 
we recognize complete agreement if we identify $\Lambda$ 
with the size of the instanton. 
The zero-size limit produces a pure-gauge configuration as it should
for self-dual $\theta^{\mu\nu}$.

Recall that the calculation of the topological charge produces a surface int%
egral at infinity, where the noncommutativity goes to zero.\footnote{Howev%
er, when $U(1)$ is gauged this is not true \cite{Furuuchi:2000vc}.} Sinc%
e our solution coincides with the standard BPST instanton in the commutative 
limit, we conclude that it has topological charge $Q=1$. This resul%
t was also obtained by Furuuchi \cite{Furuuchi:2001dx} in a direct evaluation
of $Q$ from the expressions~(\ref{curv}).


\section{Instantons from the splitting approach}

In the appendix we give a brief review of the geometric picture of the li%
near system (\ref{ls1}) in the commutative case in terms of holomorphic v%
ector bundles over the twistor space $\mathcal{P}=\mathbb{R}^4\times\mathbb{%
CP}^1$ for the space $\mathbb{R}^4$ \cite{Atiyah:1977pw,WardWells}. 
This eventually results in the Ward correspondence. For the noncommutative 
extension, the notion of vector bundles does not exist an%
ymore and is replaced by the notion of projective modules (see e.\,g.\ \cite{%
Nekrasov:2000ih}, \cite{Douglas:2001ba} and \cite{Konechny:2001wz}). Still,
the assertion of Ward's theorem remains the same on noncommutative 
$\mathbb{R}^4_\theta\times\mathbb{CP}^1$ in the following sense. 
Let $U_\pm$ be the two canonical coordinate patches covering 
$\mathbb{CP}^1$.  Taking an operator-valued %
 holomorphic matrix $f_{+-}$ restricted to $U_+\cap U_-\subset\mathbb{CP}^1$,
one may try to split $f_{+-}=\psi^{-1}_+\psi_-$ into $\psi%
_+$ and $\psi_-$ having no singularities in the spectral parameter $\lambda$ 
on $U_+$ and $U_-$, respectively. If successful then one can use the equations
\begin{equation}\label{eqfa1}
 {A}_{\bar{z}^a}\=\left.\psi_+\partial_{\bar{z}^a}\psi^{-1}_+\right|%
_{\lambda=0}\qquad\text{for}\quad a=1,2
\end{equation}
to find a self-dual gauge potential.\footnote{These equations are derived %
 and discussed in the appendix.} This defines a parametric Riemann-Hilber%
t problem. For a more detailed discussion we refer to \cite{Kapustin:2000ek%
}, \cite{Takasaki:2000vs} and \cite{Lechtenfeld:2001ie}.%

The linear system (\ref{ls1}) is equivalent to the equations$^6$
\begin{subequations}\label{cooleq1}
\begin{eqnarray}
 \psi_+(\partial_{\bar{z}^1}-\lambda\partial_{z^2})\psi^{-1}_+ \=
 \psi_-(\partial_{\bar{z}^1}-\lambda\partial_{z^2})\psi^{-1}_- &=& 
 {A}_{\bar{z}^1}-\lambda {A}_{z^2}\ ,\\
 \psi_+(\partial_{\bar{z}^2}+\lambda\partial_{z^1})\psi^{-1}_+ \=
 \psi_-(\partial_{\bar{z}^2}+\lambda\partial_{z^1})\psi^{-1}_- &=&
 {A}_{\bar{z}^2}+\lambda {A}_{z^1}\ ,
\end{eqnarray}
\end{subequations}
since using the solution (\ref{solpsi}) for $\psi$  one can construct a ma%
trix $\Upsilon$ which depends meromorphically on 
\begin{equation}
w^1 \= z^1-\lambda\bar{z}^2\ ,\qquad w^2 \= z^2+\lambda\bar{z}^1\ ,
\qquad\text{and}\quad w^3\=\lambda
\end{equation}
and acts by right multiplication so that 
\begin{equation}
\psi_+\=\psi\,\Upsilon\ .
\end{equation}
We then define $\psi_-$ by $\psi%
_-(\lambda):=[\psi^{-1}_+(-1/\bar{\lambda})]^\dagger$. In the commutati%
ve case this was shown by Crane \cite{Crane:im}. We shall now generalize %
his construction to the noncommutative setup.

\paragraph{One-instanton solution.} 
For notational simplicity let us rewrite the solution (\ref{solpsi}) 
for $\psi$ as
\begin{equation}
 \psi\=X+\lambda Y^\dagger+\frac{1}{\lambda}Y\ ,
\end{equation}
i.\,e.\ we abbreviate the known expressions by $X=G(1{+}2H)$ and 
$Y=GS=GT_2\frac{1}{K}T_1^\dagger$, where the matrices $G$, $H$, $T_i$ and $K$
are given in the previous section. 
Moreover, we expand $\psi_+$ and $\Upsilon$ 
on $U_+\cap U_-$ as
\begin{equation}
 \psi_+(x,\lambda)\=\sum_{n\in\mathbb{Z}_+}\psi_{+n}(x)\,\lambda^n
 \qquad\text{and}\qquad
 \Upsilon(w^1,w^2,\lambda)\=\sum_{n\in\mathbb{Z}}\Upsilon_n(w^1,w^2)\,\lambda^n
 \ ,
\end{equation}
since $w^3=\lambda$. Comparing coefficients of $\lambda^n$ in
$\psi_+=\psi\,\Upsilon$ then yields the equations
\begin{subequations}
\begin{eqnarray}
 \psi_{+n} &=& X\Upsilon_n+Y^\dagger\Upsilon_{n-1}+Y\Upsilon_{n+1}\qquad%
\text{for}\qquad n\geq 0\ ,\label{bla1}\\
 0         &=& X\Upsilon_n+Y^\dagger\Upsilon_{n-1}+Y\Upsilon_{n+1}\qquad%
\text{for}\qquad n< 0\ ,\label{bla2}
\end{eqnarray}
\end{subequations}
while the holomorphicity conditions $\partial_{\bar{w}^a}\Upsilon=0$ im%
ply the recursion relations
\begin{equation}\label{con}
 \partial_{\bar{z}^1}\Upsilon_n\=\partial_{z^2}\Upsilon_{n-1}\qquad\text%
{and}\qquad
 \partial_{\bar{z}^2}\Upsilon_n\=-\partial_{z^1}\Upsilon_{n-1}\ .
\end{equation}

Bearing in mind the commutative limit \cite{Crane:im} we would like to
truncate to
\begin{equation}
 \psi_+(\lambda)\=\psi_{+0}+\psi_{+1}\lambda
 \qquad\text{and}\qquad
 \Upsilon(\lambda)\=\Upsilon_{-1}\lambda^{-1}+\Upsilon_{0}+\Upsilon_{1}\lambda
 \ .
\end{equation}
We claim that it is indeed consistent to require 
$\Upsilon_{n<-1}=\Upsilon_{n>1}=0$ and $\psi_{+n<0}=\psi_{+n>1}=0$,
which reduces the infinite set (\ref{bla1}) and (\ref{bla2}) to
\begin{subequations}
\begin{eqnarray}
         0 &=& Y\Upsilon_{-1}\ ,\label{fina}\\
         0 &=& X\Upsilon_{-1}+Y\Upsilon_0\ ,\label{finb}\\
 \psi_{+0} &=& X\Upsilon_0+Y^\dagger\Upsilon_{-1}+Y\Upsilon_1\ ,\label{finc}\\
 \psi_{+1} &=& X\Upsilon_1+Y^\dagger\Upsilon_0\ ,\label{find}\\
         0 &=& Y^\dagger\Upsilon_1\ .\label{fine}
\end{eqnarray}
\end{subequations}
The truncated recursion relations (\ref{con}) imply that
\begin{subequations}
\begin{eqnarray}
\partial_{\bar{z}^a} \Upsilon_{-1} &=& 0 \= 
\partial_{{z}^a}\partial_{{z}^b}\partial_{{z}^c} \Upsilon_{-1}\ ,\\
\partial_{\bar{z}^a}\partial_{\bar{z}^b} \Upsilon_0 &=& 0 \=
\partial_{{z}^a}\partial_{{z}^b} \Upsilon_0 \qquad\text{and}\qquad
\square\,\Upsilon_0\=0\ ,\\
\partial_{\bar{z}^a}\partial_{\bar{z}^b}\partial_{\bar{z}^c}\Upsilon_1 &=& 0\=
\partial_{{z}^a} \Upsilon_1\ , 
\end{eqnarray}
\end{subequations}
so that the $2{\times}2$ matrices $\Upsilon_0$ and $\Upsilon_\pm$ 
are quadratic functions of $\{z^a\}$ and $\{\bar{z}^a\}$.
Demanding invariance under reflection on the origin and additionally imposing
$\Upsilon(\lambda)=[\Upsilon(-1/\bar{\lambda})]^\dagger$ 
the functional dependence takes the form
\begin{subequations}
\begin{eqnarray}
\Upsilon_{-1} &=& -\tau_-(z^1)^2 + \tau_+ (z^2)^2 - \tau_3\,z^1z^2\ ,\\
\Upsilon_0    &=& -\tau_3\,(z^1\bar{z}^1-z^2\bar{z}^2)
                  +2\tau_-z^1\bar{z}^2+2\tau_+z^2\bar{z}^1-\tau_4\ ,\\ 
\Upsilon_1    &=&  \tau_+(\bar{z}^1)^2 - \tau_- (\bar{z}^2)^2 
                  +\tau_3\,\bar{z}^1\bar{z}^2\ ,
\end{eqnarray}
\end{subequations}
with constant matrices $\tau_-=\tau_+^\dagger$ as well as
$\tau_3$ and $\tau_4$ (the latter two being Hermitian).

Equations (\ref{fina}) and (\ref{fine}) with 
$\Upsilon_{-1}^\dagger=-\Upsilon_1$ are clearly solved by putting
\begin{equation}
\Upsilon_{-1}\=-\sfrac{1}{2\Lambda}T_2T_1^\dagger \qquad\text{and}\qquad
\Upsilon_1   \= \sfrac{1}{2\Lambda}T_1T_2^\dagger\ ,
\end{equation}
where a convenient normalization has been chosen.
This fixes $\tau_\pm=\sfrac{1}{2\Lambda}(\sigma_1\pm\ii\sigma_2)$ and
$\tau_3=\sfrac{1}{2\Lambda}\sigma_3$, where $\sigma_i$ denotes
the Pauli matrices. The remaining condition (\ref{finb}) then
determines the matrix $\tau_4={\bf1}$. The equations (\ref{finc})
and (\ref{find}) finally serve to compute $\psi_+$.
Expressed in terms of $\{w^a\}$ coordinates, we thus arrive at
\begin{equation}
 \Upsilon(w^1,w^2,\lambda)\=\frac{1}{2\Lambda^2}
                    \begin{pmatrix} {\scriptstyle -2\Lambda^2-}\frac{w^1w%
^2}{\lambda} & \frac{(w^2)^2}{\lambda}\\[8pt]
                                    {\scriptstyle -}\frac{(w^1)^2}{\lambda} &
 {\scriptstyle -2\Lambda^2+}\frac{w^1w^2}{\lambda}
                    \end{pmatrix}\ .
\end{equation}
The matrix $\psi_+$ is then given by
\begin{eqnarray}
 \psi_+(x,\lambda) &=& \psi(x,\lambda)\, \Upsilon(w^1,w^2,\lambda)\notag\\
                 &=& -\frac{1}{\Lambda}\begin{pmatrix} \frac{1}{\sqrt{r%
^2+\Lambda^2-2\theta}} & 0\\
                                                                    0 & 
\frac{1}{\sqrt{r^2+\Lambda^2+2\theta}}
                                                    \end{pmatrix}
                 \begin{pmatrix} \bar{z}^1z^1+\Lambda^2-\lambda \bar{z}^1%
\bar{z}^2 & -\bar{z}^1z^2-\lambda(\bar{z}^1)^2\\[8pt]
                                 -z^1\bar{z}^2+\lambda(\bar{z}^2)^2 & 
\bar{z}^2z^2+\Lambda^2+\lambda\bar{z}^1\bar{z}^2
                 \end{pmatrix}
\end{eqnarray}
which coincides with Crane's solution \cite{Crane:im} in the commutative l%
imit. The matrix $f_{+-}$ then becomes
\begin{equation}
 f_{+-}(w^a,\lambda)\=\psi^{-1}_+(w^a,\lambda)\,\psi_-(w^a,\lambda)
\=\Upsilon^{-2}(w^a,\lambda)\=\frac{1}{\Lambda^2}
   \begin{pmatrix} {\scriptstyle \Lambda^2-}\frac{w^1w^2}{\lambda} & 
                   \frac{(w^2)^2}{\lambda}\\[8pt]
                   {\scriptstyle -}\frac{(w^1)^2}{\lambda} & 
                   {\scriptstyle \Lambda^2+}\frac{w^1w^2}{\lambda}
   \end{pmatrix}\ .
\end{equation}

In order to construct the gauge potential 
we need the inverse of $\psi_+$ at $\lambda=0$ which is
\begin{equation}
 \left.\psi^{-1}_+\right|_{\lambda=0}\=-\frac{1}{\Lambda}\begin{pmatrix}
 \frac{1}{\sqrt{r^2+\Lambda^2-2\theta}} & 0\\
                0 & \frac{1}{\sqrt{r^2+\Lambda^2+2\theta}}
\end{pmatrix}
                 \begin{pmatrix} \bar{z}^2z^2+\Lambda^2-2\theta & \bar{z}%
^1z^2\\[8pt]
                              z^1\bar{z}^2 & \bar{z}^1z^1+\Lambda^2+2\theta
                 \end{pmatrix}\ .
\end{equation}
With these ingredients we can use equation (\ref{eqfa1}) to reconstruct the 
gauge potential %
and hence, the curvature. What we find in this way is, of course, identical
to the result of the previous section, namely (\ref{gauge}) and (\ref{curv}).


\section{Concluding remarks}

In this paper we have extended the self-dual BPST instanton solution to Y%
ang-Mills theory defined on a noncommutative Euclidean space. We %
have chosen the noncommutative deformation matrix $\theta^{\mu\nu}$ to be %
self-dual as well, because self-dual instantons on an anti-self-dual ba%
ckground cannot be captured with our noncommutative extension %
of the Belavin-Zakharov ansatz. Our calculations demonstrate that %
noncommutativity with a self-dual $\theta^{\mu\nu}$ causes no difficult%
ies in constructing solutions. 
Potential singularities like $(r^2{-}2\theta)^{-1}$,
as occurring for the noncommutative 't Hooft instanton 
\cite{Correa:2001wv,Lechtenfeld:2001ie},
are regulated in our case by the instanton size.
Moreover, in the framework of the dressing and splitting approaches 
described in this paper we were able to solve t%
he reality problem of the gauge field which was encountered by the author%
s of \cite{Correa:2001wv} in generalizing the BPST ansatz \cite{Belavin:fg}.

It is tempting to recycle the constructed one-instanton solution as the seed 
solution in the dressing method, in order to generate multi-instantons. 
Perhaps a combination of the dressing and splitting method will do the job. 
One may hope that computing the two-instanton\footnote{
Using the dressing method Belavin and Zakharov found the two-instanton 
solution in the commutative case \cite{Belavin:cz}.} 
configuration in terms of a matrix $\Upsilon_{n=2}$ will point towards a 
recursive procedure for the construction of $n$-instantons. 
However, further work in this direction needs to be done.

The splitting and dressing approaches presented here have recently been
lifted from gauge field theory to string field theory 
\cite{Lechtenfeld:2002cu,Kling:2002ht}.
Since $10d$ superstrings in Berkovits' nonpolynomial formulation 
\cite{Berkovits:1995ab} 
as well as $4d$ self-dual strings \`a la Berkovits and Siegel
\cite{Berkovits:1997pq}
are classically integrable \cite{Lechtenfeld:2000qj},
their field equations can be linearized and classical backgrounds
can be constructed using these methods. 
Moreover, string field theory may be viewed as an infinite-dimensional
noncommutative field theory, so that the techniques of the present paper
are directly applicable. A program in this direction has been initiated
\cite{Lechtenfeld:2002cu,Kling:2002ht}.


\subsection*{Acknowledgements}

The authors are grateful to A.~Popov for useful comments and reading the 
manuscript. M.~W. thanks S.~Uhlmann and M.~Ihl for useful discussions. 
He is also grateful to the Studienstiftung des deutschen Volkes for
financial support. Z.~H. is grateful for the kind hospitality during his stay 
at Hannover University. This work was partially supported by DFG grant 
Le 838/7-2 in the priority program ``String Theory'' (SPP 1096).


\setcounter{section}{0}
\renewcommand{\thesection}{\Alph{section}}
\section{Geometry of the twistor space for $\mathbb{R}^4$}

To motivate the linear system (\ref{ls1}) and to understand the geometry %
behind it, this appendix analyzes its commutative analog. 
It has a geometri%
cal interpretation in terms of holomorphic bundles over the twistor space %
 $\mathcal{P}=\mathbb{R}^4\times\mathbb{CP}^1$ for the space $\mathbb{R%
}^4$ \cite{Atiyah:1977pw,WardWells}.%

It is well known that the two-sphere $S^2\cong\mathbb{CP}^1$ %
 can be covered by two coordinate patches $U_\pm$ and hence, also the whole tw%
istor space $\mathcal{P}$: 
\begin{equation}
 \mathcal{P}\=\mathcal{U}_+\,\cup\mathcal{U}_-\ ,\qquad \mathcal{U}_+\=
\mathbb{R}^4\times U_+\ ,\qquad\mathcal{U}_-\=\mathbb{R}^4\times U_-\ ,
\end{equation}
with $U_+=\mathbb{CP}^1\setminus\{\infty\}$ and $U_-=\mathbb{CP}^1
\setminus\{0\}$. The complexified tangent space $\mathbb{C}^4\cong\mathbb{R}%
^4\otimes\mathbb{C}$ of $\mathbb{R}^4$ can be decomposed into two subspac%
es with respect to a chosen constant almost complex structure 
$J=(J^\mu_{\,\,\nu})$ on 
$\mathbb{R}^4$, i.\,e.\ $\mathbb{C}^4\cong\mathcal{V}\oplus\bar{\mathcal{V}}$ %
with$^8$
\begin{equation}
 \mathcal{V}\=\{V\in\mathbb{C}^4\,|\,J^\mu_{\,\,\nu}V^\nu=\ii V^\mu\}\qquad
\text{and}\qquad%
 \bar{\mathcal{V}}\=\{V\in\mathbb{C}^4\,|\,J^\mu_{\,\,\nu}V^\nu=-\ii V^\mu\}\ .
\end{equation}
For instance, on $\bar{\mathcal{V}}$ we may take the basis
\begin{equation}
 (\bar{V}_1^\mu) \= \left({\textstyle\frac{1}{2},\frac{\ii}{2},-\frac{1}{2%
}\lambda,-\frac{\ii}{2}\lambda}\right)
                        \qquad \text{and} \qquad
 (\bar{V}_2^\mu) \= \left({\textstyle\frac{1}{2}\lambda,-\frac{\ii}{2}
\lambda,\frac{1}{2},-\frac{\ii}{2}}\right)\ ,
\end{equation}
where $\lambda$ is the local holomorphic coordinate on the two-sphere $S^%
2\cong SO(4)/U(2)$ which parametrizes $J$, i.\,e.\ $\lambda=(\xi^1+\ii\xi^2)%
/(1+\xi^3)$ with $\xi^a\xi^a=1$. Therefore, we can introduce 
on $\mathcal{U}_+$ (and similarly on $\mathcal{U}_-$) the anti-ho%
lomorphic vector fields $\bar{V}_a=\bar{V}_a^\mu\partial_\mu$ ($a=1,2%
,3$)\footnote{For a detailed description see e.\,g.\ \cite{Popov:1998pc}.}
\begin{equation}
 \bar{V}_1\=\partial_{\bar{z}^1}-\lambda\partial_{z^2}\ ,\qquad\bar{V}_2\=%
\partial_{\bar{z}^2}+\lambda\partial_{z^2}\qquad\text{and}\qquad
 \bar{V}_3\=\partial_{\bar{\lambda}}\ ,
\end{equation}
where we have chosen the standard complex structure on $S^2\cong\mathbb{C%
P}^1$. This means that the coordinates (\ref{cc}) are the canonical compl%
ex coordinates on $\mathbb{R}^4\cong\mathbb{C}^2$ corresponding to $\lambda=0$.
Hence, the appropriate coordinates on the twistor space $\mathcal{P}$ are
\begin{subequations}
\begin{eqnarray}
 w^1=z^1-\lambda\bar{z}^2\ ,\qquad w^2=z^2+\lambda\bar{z}^1\qquad &
\text{and}\qquad & w^3=\lambda\qquad\text{on}\quad\mathcal{U}_+\ ,\\
 \tilde{w}^1=\tilde{\lambda} z^1-\bar{z}^2\ ,\qquad \tilde{w}^2=\tilde{%
\lambda} z^2+\bar{z}^1\qquad &\text{and}\qquad &%
 \tilde{w}^3=\tilde{\lambda}\qquad\text{on}\quad\mathcal{U}_-\ ,
\end{eqnarray}
\end{subequations}
which are related on the intersection $\mathcal{U}_+\cap\mathcal{U}_-
\cong\mathbb{R}^4\times\mathbb{C}^*$ via
\begin{equation}
 w^1=\frac{\tilde{w}^1}{\tilde{w}^3}\ ,\qquad 
 w^2=\frac{\tilde{w}^2}{\tilde{w}^3}
 \qquad\text{and}\qquad w^3=\frac{1}{\tilde{w}^3}\ .
\end{equation}
Moreover, we can construct a local basis of 1-forms $\bar{\theta}^a(\bar{V%
}_b)=\delta^a_{\,\,b}$ on $\mathcal{U}_+$ which read
\begin{equation}
 \bar{\theta}^1 \= \gamma\,(\D{}\bar{z}^1-\bar{\lambda}\D{}z^2)\ ,\qquad
 \bar{\theta}^2 \= \gamma\,(\D{}\bar{z}^2+\bar{\lambda}\D{}z^1)
 \qquad\text{and}\qquad  \bar{\theta}^3 \= \D{}\bar{\lambda}\ ,
\end{equation}
with $\gamma=(1+\lambda\bar{\lambda})^{-1}$.

On the principal bundle $P=\mathbb{R}^4\times U(2)$ over $\mathbb{R}^4$ %
 the connection $1$-form ${A}$ determines the connection $D=\D{%
}+{A}$ on $P$. Furthermore, let $\rho\,:\,U(2)\rightarrow GL(2,\mathbb{C})$ 
be the fundamental representation of $U(2)$. Then, as usual, w%
e associate to $P$ the complex vector bundle $E=P\times_\rho\mathbb{C}^%
2$. Using the projection $\pi\,:\,\mathcal{P}\rightarrow\mathbb{R}^4$ we %
can pull back $E$ to a bundle $\pi^*E$ over $\mathcal{P}$. By definition, %
 the pulled-back connection $1$-form $\pi^*{A}$ on $\pi^*E$ is fl%
at along the fibers $\mathbb{CP}^1$ and, hence, the pulled-back connection
$\pi^*D$ is nothing but $\pi^*D|_{\mathcal{U}_+}=D+\D{}\lambda\,
\partial_\lambda+\D{}\bar{\lambda}\,\partial_{\bar{\lambda}}$. Writing $\pi^*%
{A}=:{B}+\bar{{B}}$ with $\bar{{B}}=:\bar{{B}}_a\bar{\theta}^a$ 
we discover the components
\begin{equation}
 \bar{{B}}_1 \= {A}_{\bar{z}^1}-\lambda{A}_{z^2}\ ,\qquad
 \bar{{B}}_2 \= {A}_{\bar{z}^2}+\lambda{A}_{z^1}
 \qquad\text{and} \qquad \bar{{B}}_3 \= 0\ ,
\end{equation}
whereby on the intersection $\mathcal{U}_+\cap\mathcal{U}_-$ we have 
$\bar{{B}}_a=\lambda\tilde{\bar{{B}}}_a$. The pulled-back c%
onnection $\pi^*D$ is then given in terms of ${B}$ by $\pi^*D=
\partial_{{B}}+\bar{\partial}_{\bar{{B}}}$ with $\bar{\partial}_%
{\bar{{B}}}=\bar{\partial}+\bar{{B}}=\bar{\theta}^a(\bar{V}_a+\bar{B}_a)$. 

Let us consider the equations 
$\bar{\partial}_{\bar{{B}}}s=0$ for a local smooth section $s$ of $\pi^*E$. %
 By definition the local solutions of these equations are just the %
 local holomorphic sections of $\pi^*E$. The bundle $E^\prime:=\pi^*E
\rightarrow\mathcal{P}$ is then termed holomorphic iff these equation%
s are compatible in the sense of $\bar{\partial}_{\bar{{B}}}^2=%
0$. Writing down $\bar{\partial}_{\bar{{B}}}s=0$ explicitly e.\,g.\ on %
$\mathcal{U}_+$ one realizes that $s_+:=s|_{\mathcal{U}_%
+}$ does not dependent on $\bar{\lambda}$. The compatibility equations $\bar{%
\partial}_{\bar{{B}}}^2=0$ coincide with the SDYM equations (\ref{eom}). 
Therefore, we have local solutions $s_\pm$ on $%
\mathcal{U}_\pm$ with $s_+=s_-$ on the intersection $\mathcal{U}_+\cap
\mathcal{U}_-$. Note that we can always decompose $s_\pm=\psi_\pm%
\eta_\pm$, where $\psi_\pm$ lie in the complexified gauge group $U(2)
\otimes\mathbb{C}\cong GL(2,\mathbb{C})$, are nonsingular on $\mathcal{U}_\pm%
$, and satisfy $\bar{\partial}_{\bar{{B}}}\psi_\pm=0$ on $\mathcal{U}_\pm$. 
The vector functions $\eta_\pm\in\mathbb{C}^2$ are holomorphi%
c on $\mathcal{U}_\pm$, i.\,e.\ they are only functions of $\{w^a\}$ and $\{%
\tilde{w}^a\}$, respectively. We therefore have
\begin{subequations}\label{cooleq}
\begin{eqnarray}
 \psi_+(\partial_{\bar{z}^1}-\lambda\partial_{z^2})\psi^{-1}_+ \=%
 \psi_-(\partial_{\bar{z}^1}-\lambda\partial_{z^2})\psi^{-1}_- &=&%
 {A}_{\bar{z}^1}-\lambda {A}_{z^2}\ ,\\[8pt]
 \psi_+(\partial_{\bar{z}^2}+\lambda\partial_{z^1})\psi^{-1}_+ \=
 \psi_-(\partial_{\bar{z}^2}+\lambda\partial_{z^1})\psi^{-1}_- &=&%
 {A}_{\bar{z}^2}+\lambda {A}_{z^1}\ ,
\end{eqnarray}
\begin{equation}
 \partial_{\bar{\lambda}}\psi_+\=\partial_{\bar{\lambda}}\psi_- \= 0
\end{equation}%
\end{subequations}
on $\mathcal{U}_+\cap\mathcal{U}_-$, and consequently
\begin{equation}\label{eqfa}
 {A}_{\bar{z}^a}\=\left.\psi_+\partial_{\bar{z}^a}\psi^{-1}_+
 \right|_{\lambda=0}\qquad\text{for}\quad a=1,2\ .
\end{equation}
Furthermore, the vector functions $\eta_\pm$ are related via 
\begin{equation}
 \eta_+\=f_{+-}\,\eta_- \qquad\text{with}\qquad f_{+-}\=\psi^{-1}_+\,\psi_-
 \qquad\text{on}\quad \mathcal{U}_+\cap\mathcal{U}_-\ ,
\end{equation}
which implies the holomorphicity of $f_{+-}$.

In summary, we have described a one-to-one correspondence between gauge e%
quivalence classes of self-dual connection $1$-forms ${A}$ on a c%
omplex vector bundle $E$ over $\mathbb{R}^4$  and equivalence classes of %
holomorphic vector bundles $E'$ over the twistor space $\mathcal{P}$ triv%
ial on $\mathbb{CP}^1\hookrightarrow\mathcal{P}$. A local gauge tra%
nsformation of the gauge potential ${A}$ is reflected by 
$\psi_\pm\mapsto\psi^g_\pm:=g^{-1}\psi_\pm$ and, hence, leaves the %
 transition function $f_{+-}$ invariant. On the other hand, the gauge pot%
ential ${A}$ is inert under a transformation $\psi_\pm%
\mapsto\psi^{h_\pm}_\pm:=\psi_\pm h_\pm^{-1}$, where $h_\pm$ live in th%
e complexified gauge group and are regular holomorphic on $\mathcal{U}_\pm$, 
respectively. This is known as the twistor correspondence or the Eucl%
idean version of Ward's theorem \cite{Atiyah:1977pw,WardWells}.


\vfill\eject

\end{document}